# WHAT IS SPECIAL ABOUT THE PLANCK MASS?

C. Sivaram

Indian Institute of Astrophysics, Bangalore 560034

**Abstract:**

Planck introduced his famous units of mass, length and time a hundred years ago. The many interesting facets of the Planck mass and length are explored. The Planck mass ubiquitously occurs in astrophysics, cosmology, quantum gravity, string theory, etc. Current aspects of its implications for unification of fundamental interactions, energy dependence of coupling constants, dark energy, etc. are discussed.

A hundred years ago, Max Planck introduced in 1906, the famous units of mass, length and time, constructed solely out of the three fundamental constants, $\hbar = h/2\pi$, c and G [1]. Here $\hbar$ is the Planck constant introduced by him in 1900 (heralding the birth of quantum physics), c is the velocity of light (governing laws of relativity) and G is the Newtonian gravitational constant. This union of the three very basic aspects of the universe (i.e., the quantum, relativistic and gravitational phenomena) is embodied in the three unique quantities:

Planck mass, $M_{pl} = \left(\dfrac{\hbar c}{G}\right)^{1/2} = 2.2 \times 10^{-5} \, gm$ ...(1)

Planck length, $L_{pl} = \left(\dfrac{\hbar G}{c^3}\right)^{1/2} = 1.6 \times 10^{-33} \, cm$ ...(2)

Planck time, $t_{pl} = \left(\dfrac{\hbar G}{c^5}\right)^{1/2} = 5 \times 10^{-44} \, \sec$ ...(3)

Equations (1)-(3) involve only, $\hbar$, c and G and do not depend upon masses associated with any particle like the electron or proton or corresponding Compton lengths. The Planck mass, for instance, seems to be a very basic unit of mass. However, it is $10^{19}$ times the proton mass! This corresponds to an energy scale of $10^{19} \, GeV$. At this energy (or mass), the gravitational coupling strength given by the dimensionless quantity $GM_{pl}^2/\hbar c$ becomes strong, i.e. ~ 1, (as compared to the electromagnetic fine structure $\alpha = \dfrac{e^2}{\hbar c} = 1/137$). For two protons, the gravitational interaction is proportional to $\sim Gm_P^2/\hbar c \sim 10^{-38}$. $m_P$ is the proton mass. We can write for the dimensionless gravitational coupling (or fine structure constant), $\alpha_g \sim Gm^2/\hbar c \sim \left(m/M_{pl}\right)^2$, showing that only for $m \sim M_{pl}$, gravity becomes strong i.e. comparable to the strength of strong interactions. The Planck length corresponds to the Compton Length corresponding to the Planck Mass. i.e. $\hbar/cM_{pl} = (\hbar G/c^3)^{1/2}$. (Given by eq. (2)).

We shall return to the significance of these quantities (eqs. (1) – (3)) later.

In this context, it must be pointed out that more than 25 years before Planck introduced these quantities, the Irish physicist Johnstone Stoney [2] in 1881 introduced the following quantities of mass, length and time based on the then (newly discovered) unit of electric (electron) charge e and the gravitational constant G. As this was in purely classical era of Maxwell (i.e., pre-quantum!), only e and G were involved, and also c of course (velocity of electromagnetic waves!).

Stoney introduced the units:

Length, $L_e = \left(\dfrac{Ge^2}{c^4}\right)^{1/2} = 10^{-35} cm$ …(4)

Time, $t_e = \left(\dfrac{Ge^2}{c^6}\right)^{1/2} = 3\times 10^{-46}$ sec …(5)

Mass, $M_e = \left(\dfrac{e^2}{G}\right)^{1/2} = 10^{-6} g$ …(6)

The quantities (4) – (6) are related to the Planck units (1) – (3) by just the factor $\sqrt{\alpha}, (\alpha = e^2/\hbar c)$, i.e. multiplying by this factor. Eq (6) would signify that for a mass $M_e$, and charge $e$, the gravitational and electromagnetic interactions become of equal strength, i.e. $GM_e^2 = e^2$.

(Or two particles of mass $M_e$ each and charge of each being $e$ would have equal balance between their electrostatic and gravitational interactions). Again, the mass $M_e$ is far larger than say, the electron mass of $10^{-27} gm$. The origin of the masses of fundamental particles (such as the electron or proton) is still a major unsolved problem in physics. Though the electric charge and spin (units of $\hbar$) are universally the same. So it would be desirable to have a mass formula just in terms of universal interaction couplings (like $e$) and basic constants like G and $\hbar$. However $M_{pl}$ and $M_e$ are considerably larger than the known masses of elementary particles. We shall return to this problem later. As we see above, $M_{pl}$ excludes $e$, while $M_e$ does not have $\hbar$! Is it possible to include all the four basic constants ($\hbar$, G, $c$ and $e$) in defining a mass? This can be done. Consider for example, the balance between the energy density of a sufficiently large magnetic flux and its self gravitational, energy density $\frac{GM^2}{8\pi R^4} = \frac{B^2}{8\pi}$, giving $GM^2 = B^2 R^4 = \Phi^2$, $\Phi = BR^2$ is the flux. As magnetic flux can be quantized in units of $(\hbar c/e)$, the above relation gives

$$M = \frac{\hbar c}{\sqrt{Ge}} = 2.5 \times 10^{-4}\, gm \qquad \ldots(7)$$

as a basic mass unit having $\hbar$, c, G and e!

We can also define the length:

$$l = \frac{\hbar \sqrt{G}}{ec} = 1.9 \times 10^{-32}\, cm,$$

And $t = \frac{\hbar \sqrt{G}}{ec^2} = 6 \times 10^{-43}\, s$.

As eq (7) involves $e$ in addition to $\hbar$, $c$ and G it may have a claim to be a more basic unit (of mass)!

However, it is remarkable that we cannot construct a dimensionless quantity involving $\hbar$, $c$, G and $e$ in three spatial dimensions!

(Like we have $\alpha = e^2/\hbar c = 1/137$, a dimensionless number in our three dimensional world).

The dimensionless quantity in '$n$' spatial dimensions (involving $\hbar$, c, G and e) is:

$$e^{n-1} \hbar^{2-n} c^{n-4} G^{(3-n)/2} = A \qquad \ldots(8)$$

For $n = 3$ (our world!)

We have $\dfrac{e^2}{\hbar c} = \alpha$

We need at least $n = 5$; to have a dimensionless quantity having all of the above including G. String theory with its six compact dimensions perhaps has a clue. In superstring theory, the particle masses (energies) are pictured as excitations of a one-dimensional object (string) of about a Planck length, this string, having a very high tension, $T_{pl} \approx c^2/G_N$, i.e. *Planck mass/ Planck length*. However as the mass $(M_{pl})$ is far larger than a typical meson or baryon mass it is not clear how one can arrive at masses corresponding to the actually observed particle mass spectrum. If one has a piano string, the frequency of the $m^{th}$ octave and the $p^{th}$ note is given by a formula like

$$f_{mp} = f_0 \times 2^m \left(\sqrt[12]{2}\right)^P \qquad \ldots(9)$$

Where, $f_0$ is the lowest (fundamental) frequency. It was suggested in ref [3,4] that a similar mass formula for the excitations of a Planck length string involving the superstring tension $T_{pl}$, the Planck mass, the electric charge etc. spectrum of particles. This formula was given of the form:

$$M = nk\left[\dfrac{M_{pl}c}{e} \dfrac{1}{2\sqrt{T_{pl}}}\right]^n m_e = nk\left[\sqrt{G_N} \dfrac{M_{pl}}{2e}\right]^n m_e \qquad \ldots(10)$$

Here $m_e$, the electron rest mass, being the lowest known mass of a charged particle takes the place of $f_0$ in equation (9). *n.k* are integers.

Indeed, the quantity within square brackets in equation (10) can be written as the ratio of the fundamental masses M (equation 5) and $M_{Pl}$! M and $M_{Pl}$ are constructed slowly out of the basic constants, $\hbar$, G, *e* and *c*. So it is remarkable if all the particle masses can arise from them. As far as the origin of the electron mass $m_e$ which enters into equation (10), we shall take up this later.

For example, *n=2, k=3*, gives the muon mass, *n=2, k=4,* gives the pion mass, *n=3, k=4* the $\Delta$ resonance, *n=3, k=6*, the D meson, *n=3, k=10*, the $\Psi$, *n=4, k=4*, the upsilon particle. Several more particle masses can be obtained. All of the above values agree to within less than *0.5%* with experiment!

The logarithmic forms of the formulae eqs. (9) and (10), i.e., for the frequency modes of the vibrating piano string and for the modes of the vibrating Planck string(!) are respectively:

$$\ln f = \ln f_0 + \ln m + \frac{p}{12} \ln 2$$

$$\ln M = \ln n + \ln k + \frac{n}{2} \ln\left(\sqrt{G_N} M_{pl} \Big/ e\right)$$

(In the string picture, the various particles are due to the vibrating modes of the Planck size string!)

The Planck mass also ubiquitously enters into various parameters in astrophysics and cosmology. For instance, the Chandrasekhar mass limit for white dwarfs can be written as:

$$M_{ch} = \left(\frac{\hbar c}{G m_P^2}\right)^{3/2} m_P = \frac{M_{pl}^3}{m_P^2} = 1.5 M_\odot ! \qquad \ldots(11)$$

($m_P$ is the proton mass).

The total number of baryons in the universe is $\sim \left(\frac{M_{pl}}{m_P}\right)^4$ and the photon – to – baryon ratio (another constant parameter) in standard cosmology can be written as

$$\frac{n_\gamma}{n_b} \sim \left(\frac{M_{pl}}{m_P}\right)^{1/2} \sim 10^{9.5} !$$

The Hubble age to the Planck time is $\sim 10^{61}$, which is the same as $\frac{M_U}{M_{pl}}$, $M_U$ is the total mass of the universe. The ratio of the large-scale structure scale $\sim M_{pc}$, to the Planck length is the same as $\left(M_{pl}/m_P\right)^3$. A dark matter particle of mass $m_D$, can give rise to a structure of mass $\sim \dfrac{M_{pl}^3}{m_D^2}$, (which for $m_D \sim 1 keV$, gives mass $\sim 10^{12} M_\odot$, a typical large galaxy mass). There are any number of such relations involving the Planck mass. [3,5,6]

The entropy of a black hole in units of $k_B$ (Boltzmann constant) is just $\left(\dfrac{M_{BH}}{M_{pl}}\right)^2$, where, $M_{BH}$ is the black hole mass. Some authors have conjectured that evaporating black holes leave a remnant of mass $\sim M_{pl}$, and that part of the DM could be primordial black holes each of mass $M_{pl}$. The smallest primordial black hole mass would be $\sim M_{pl}$; created at Planck epoch $\left(\sim t_{pl}\right)$ in the early universe.

Of course we also know that there are weak interactions, characterized by the universal Fermi constant $G_F \left(= 1.5 \times 10^{-49} \, erg cm^3\right)$ and the strong interaction (quark-gluon etc.). The Planck mass, despite its basic nature, does not involve $G_F$ or $e$, the electric charge. The Fermi constant, being dimensional like Newton's gravitational constant $G$, also gives rise to a characteristic length, the beta-decay length, given by

$$L_W = \left(\dfrac{G_F}{\hbar c}\right)^{1/2} = 7 \times 10^{-17} \, cms$$

With a corresponding mass $M_W = \hbar/L_W c \approx 240 GeV$.

In this context, it is also interesting that we can construct a mass out of $e, G_F, G, \hbar, c$, etc. This takes the form:

$$m = \dfrac{\hbar^3 \sqrt{G}}{e G_F c} \qquad \ldots(12)$$

A possible derivation is from particle magnetic moments. In the Klein-Kazula picture, a Dirac particle acquires a magnetic moment (induced by gravity) [7] of order $\hbar\sqrt{G}/c$.

In electroweak theory, a neutrino of mass $m_\nu$ acquires a magnetic moment due to radiative connections and this given by: [8]

$$\mu_\nu \approx \frac{eG_F m_\nu}{\hbar^2}$$

(As the gravity induced term would be the smallest possible, equating the above expressions would give a lower limit to the neutrino mass) i.e. given by:

$$m_\nu = \frac{\hbar^3 \sqrt{G}}{eG_F c}, \text{ Which is given by equation (12)!}$$

This works out to $m_\nu \approx 1.2 \times 10^{-37} g \approx 7 \times 10^{-5} eV$ !

So this shows that all masses constructed solely from basic interaction coupling constants need not be large like the Planck mass!

Eq (12) has the couplings of gravity, weak and electromagnetic interactions (apart from $\hbar$ and c). So this value of $m_\nu$ should have some basic significance! Neutrino oscillation experiments do imply a $\Delta m^2 \sim 10^{-4} eV^2$.

As neutrinos constitute part of the dark matter (DM) in the universe, the recent WMAP limit [9] on the upper bound of $0.7 eV$ for the neutrino mass could be relevant.
We will discuss this soon. Currently, WMAP, along with the results of the supernova cosmology Project has provided strong evidence that the universe is dominated by Dark Energy (DE) that constitutes at least 70 percent (of the universe) of the inventory. [10]

There is further strong evidence [11], from the Supernova Legacy survey etc. that the DE may just be Einstein's cosmological constant $\Lambda$.

The cosmological constant introduces a curvature or length scale. This can be associated with a particle mass given by [12]

$$m = \frac{\hbar\sqrt{\Lambda}}{c} \qquad \ldots(13)$$

For the observed value of $\Lambda(\sim 10^{-56} cm^{-2})$, this gives: $m \approx 3\times 10^{-66} gm$. If there is coupling between DE and DM, and considering that the neutrino takes part in both weak and gravitational interactions, we can modify equation (13) into a neutrino mass given by: [13]

$$m_\nu = \frac{c\sqrt{\Lambda}}{\hbar} \frac{G_F}{G} \frac{\Omega_{DM}}{\Omega_{DE}} = 0.7 eV \qquad \ldots(14)$$

Where $\Omega_{DM}/\Omega_{DE}$ is the ratio of dark matter to dark energy. Eq (14) is just the bound given by WMAP. [10]

Following the approach of Hayakawa [14], for example, and the present author [15], it is interesting that one can give a Machian type picture for the neutrino mass. If $g_W$ is the typical weak charge, $l_W$ the beta decay length $\sim \left(G_F/\hbar c\right)^{1/2}$, $N_\nu$ the total number of neutrinos in the Universe then the fluctuation in the total weak interaction energy (of all the $N_\nu$ neutrinos in the universe which are distributed over a region of radius $\Lambda^{-1/2}$) gives rise to the local weak interaction energy $\sim g_W^2/l_W$.

I.e., $\sqrt{N_\nu} g_W^2 \sqrt{\Lambda} \approx g_W^2/l_W$

$$l_W = \left(\frac{G_F}{\hbar c}\right)^{1/2} = \frac{1}{\sqrt{N_\nu \Lambda}} \qquad \ldots(15)$$

This gives the Fermi constant as:

$$G_F = \frac{\hbar c}{N_\nu \Lambda} \qquad \ldots(16)$$

Where, $\Lambda$ is the cosmological constant.

Equation (16) gives the total number of neutrinos as:

$$N_\nu = \frac{\hbar c}{G_F \Lambda} \approx 2 \times 10^{88} \qquad \ldots(17)$$

(For an observed dark energy corresponding to $\Lambda \sim 10^{-56} cm^{-2}$)

For a universe volume, $\sim 2\pi^2 R_H^3 \sim 2\pi^2 \Lambda^{-3/2} \sim 2 \times 10^{85} cc$, this gives a total neutrino density (at present epoch) $\approx 10^3 cm^{-3}$, precisely what is expected.

(For a predicted background neutrino temperature of $2.1K = \left(4/11\right)^{1/3} T_\gamma$, $T_\gamma = 2.7K$, $n_\nu$ (number density) for six species of $\nu's$ (three flavours + antiparticles) is just thousand!)

For above case (from equation 17):

$$n_\nu = \frac{\hbar c \sqrt{\Lambda}}{2\pi^2 G_F} \qquad \ldots(18)$$

Considering dark energy density $\sim \Lambda c^2 / 8\pi G$ and with the ratio of $\Omega_{DM}/\Omega_{DE}$, gives precisely the formula given by equation (14) for the upper limit of the neutrino mass.

Hayakawa [14] related the number of electrons (protons) in the universe to the electron radius. As the number of electrons is $10^{10}$ times lesser than $N_\nu$, $\left(N_e \approx 10^{78}\right)$ equation (15) would give the classical electron radius ($\sim 5 \times 10^{-13} cm$) for the same $\Lambda$.

The electron mass itself can be related to the dark energy [13,6] given by $\Lambda$. By considering a wave packet of spread $r$, and equating its gravitational self energy to the repulsive DE density of the cosmological vacuum gave, $r^6 \approx \frac{L_{pl}^4}{\Lambda}$, or $r = \frac{L_{pl}^{2/3}}{\Lambda^{1/6}} = \frac{1}{\Lambda_{pl}^{1/3} \Lambda^{1/6}}$.

($\Lambda_{pl} = L_{pl}^{-2}$ is the Planck curvature)

If this localized packet acquires an electron charge, then its electrostatic self energy just gives the electron rest mass $m_e$ as:

$$m_e = \frac{\alpha \hbar}{c} \Lambda_{pl}^{1/3} \Lambda^{1/6} \qquad \ldots(19)$$

In the mass formula given above, the only 'assumed' mass was $m_e$. So we can give a mass formula (which gives the masses of a large number of known particles) entirely in terms of $e, G, \hbar, c$ as: (and a constant cosmic parameter $\Lambda$, characterising the dominating DE)

$$M = nk\left[\sqrt{G_N} \frac{M_{pl}}{e} \frac{1}{2}\right]^n \frac{\alpha \hbar}{c} \Lambda_{pl}^{1/3} \Lambda^{1/6} \qquad \ldots(20)$$

It is often felt [16] that unlike other units (i.e. length and time), the standard of mass, 1 kilogram, is not expressible in fundamental constants. It is only compared with a prototype cylinder kept in France.

A combination of coupling constants of all the four basic interactions (including the strong interaction gluon coupling $\alpha_S = 0.13$), which along with $N = 10^{78}$ (total number of baryons) gives close to one kilogram is, see also [17] is:

$$\frac{\hbar^3 \sqrt{G}}{e G_F c} \frac{\sqrt{N}}{\alpha_S} = 1 Kg \qquad \ldots(21)$$

If we do not wish to treat $\sqrt{N}$ as a separate quantity, it can be related to equation (18) through the photon to baryon ratio given by

$$\left(\hbar c / G m_P m_e\right)^{1/4} \sim \left(\frac{M_{pl}^2}{m_P m_e}\right)^{1/4} = n_\gamma / n_B .$$

Thus, $N_\nu \left(\frac{m_P m_e}{M_{pl}^2}\right)^{1/4} = \left(\frac{\hbar c}{G_F \Lambda}\right) \left(\frac{m_P m_e}{M_{pl}^2}\right)^{1/4} = 10^{78} \qquad \ldots(22)$

One can substitute for $m_e$ from equation (19) and $m_P$ from ratio of $m_P/m_e$ given in [3], to get the formula given by equation (21) entirely in terms of, $e, G, \hbar, c\, \&\, \Lambda$, which appear to be universal unchanging constants. The precise value of these constants, then fix the standard of mass.

The problem of understanding the origin of mass (which is still unsolved) was initiated by Planck's construction of such a quantity solely in terms of the basic physical constants, $\hbar, c\, \&\, G$ and without involving the mass of any other particle. However the Planck mass is too large as compared to the familiar elementary particles. As we have shown, it is possible to construct, in the spirit of Planck, masses constructed purely out of the coupling constants of the four basic interactions (and perhaps a basic parameter like $\Lambda$) and that these masses correspond to those of the familiar elementary particles. The Planck mass plays a basic role in determining these masses.

## Appendix

**1. Planck mass as a limiting energy**

In many discussions of the highest possible energies, the Planck energy plays the role of an 'ultra-violet' cut off. At that energy, $E_{pl}$, the particle Compton length, i.e., $\hbar c / E_{pl}$, equals its gravitational radius, i.e., $GE_{pl}/c^4$, so that the particle is trapped in its own gravitational field. Any particle or quantum of radiation with that energy, $(\sim E_{pl})$ would have a wavelength $\lambda \sim L_{pl}$, and its gravitational self energy would equal its energy, i.e., $\hbar c/\lambda \sim Gm^2/\lambda$, which gives a definition of the Planck mass! This implies an upper limit to cosmic ray particle energies $\sim 10^{23}\,eV$, few orders larger than the highest energies seen so far $\sim 10^{21}\,eV$. This also tells us that virtual particles in QED and other field theories cannot have arbitrary high energies so that the particle (eg. electron) self energies get truncated to the form $\alpha \sim \ln(\hbar c / Gm_e^2)$, explaining the smallness of $G$ relative to $\alpha$ !

The Planck length corresponding to a maximal curvature of $R_{max} \approx c^3/\hbar G \sim 10^{66} cm^{-2}$, in turn imposes the ultimate quantum limit on geometrical measurements giving the smallest spatial resolution $\sim L_{pl}$. The maximal curvature also implies a maximal field strength $\sim c^{7/2}/(\hbar G)^{1/2}$ [18]. The Planck units physically imply upper bounds or limits (dictated by unity of quantum physics with general relativity) on several properties of matter such as density, elastic strength, temperature $(\sim 10^{32} K)$, etc., as well as absolute upper limits on computational and information processing rates $(\sim 10^{44} bits/\sec)$ and power generation $(\sim 10^{59} W)$ etc. These have also significance for the very earliest epochs of the universe (i.e. the so called Planck era given by $t_{pl}$, eq. (3)). The 'smooth' structure of space-time manifolds could undergo drastic changes at the Planck scale, perhaps giving rise to discrete structure of space-time, where even spatial co-ordinates become non-commutative. I.e. $[x, y] = iL_{pl}^2$, apart from $[x, p_x] = i\hbar$, etc.

These quantum fluctuations in space-time, could have microscopic consequences which in principle could be detected in deviations from Newton's gravity law, corrections in atomic spectroscopy, with consequences for gravitational wave detectors (in their displacement noise spectrum), k-meson decays, time delays in gamma ray bursts, etc. For a review of many of these effects see for eg. [18,19]. At around the Planck scale, one expects a modification of the uncertainty principle (giving a so called generalised uncertainty principle GUP). A typical GUP relation (also suggested in superstring theories) is [20]:

$$\Delta x \Delta p \geq \hbar + \frac{\lambda}{\hbar}(\Delta p)^2 \qquad \ldots(23)$$

Where, $\lambda$ has the dimensions of $L_{pl}^2$ (i.e., Planck area, in loop quantum gravity models area is quantized in units of $L_{pl}^2 \approx 10^{-66} cm^2 \approx 1$ atto shed; 1 shed $= 10^{-24}$ barns).

The above relation implies a minimal length $\sim 2\sqrt{\lambda}$. In string theories, $\lambda \sim L_S$, where $L_S$ is the string length and implies a minimal length and non-divergent self energies.

The usual phase space giving the total number of quantum states is modified from $\frac{dVd^3p}{(2\pi\hbar)^3}(\Delta x \Delta p \sim 2\pi\hbar)$, to, $\frac{dVd^3p}{(2\pi\hbar)^3(1+\lambda p^2)^3}$, $p^2 = p_i p^i$, $i = 1,2,3$.

The non-commutativity $[x, p]$ becomes:

$[\hat{x}, \hat{p}] = i\hbar(1+\lambda p^2)$, which can be interpreted as a 'modification' of $\hbar$ to $\hbar' = \hbar(1+\lambda p^2)$ (this was first suggested in ref [6]).

The above relations have interesting consequences for the early universe and for deforming the Planck spectrum in black holes at close to the Planck temperature. [20,21].

## 2. Variation of G and the Planck scale

In defining the Planck energy scale etc. we have used the Newtonian coupling G, which is the value for the macroscopic gravitational interactions at 'low' energies. If in analogy with other interactions the gravitational coupling varies with energy at high energies, i.e., $G = G(E)$.

For abelian fields like electromagnetism, the coupling rises logarithmically with energy, so that the value of $\alpha$ at the Z-boson mass (i.e. $\sim 90 GeV$) is not $1/137$ but $1/128$!

For strong interactions, coupling fall with energy E:

$\alpha_S(E) = \frac{\alpha_S(0)}{\ln\left(\frac{E}{\Lambda_S}\right)}$, $\Lambda_S$ is the QCD scale.

As is well known especially in the minimal supersymmetric standard model couplings of electroweak and strong interaction become same at a high energy scale $\sim 10^{16} GeV$, at which energy, gravity has a relative strength of $\sim 10^{-6}$. The gravitational dimensionless coupling scales as $\sim \frac{GE^2}{\hbar c}$, so is expected to continue rising with energy till it becomes 'strong' (~1) at the Planck energy! However if $G$ itself scales with energy by analogy with other interactions, for instance the weak interaction with strength given by the Fermi coupling constant $G_F$ is comparatively 'weak' as $G_F$ is related to the large mass of the intermediate boson $m_W \sim 10^2 \, GeV$ i.e. $G_F \propto 1/m_W^2$. [22]

For gravity $G \propto 1/m_{pl}^2$. So as the Planck mass is large, gravity is weak. All quantum gravitational processes become 'strong' at the Planck energy. However if $G$ changes with time in the early universe (eg. $G \propto t^{-1}$) or with energy $G \propto E^{-2}$, or the gravity coupling scales logarithmically with energy (like other interactions) we can no longer define the Planck units with constant values at high energies. So the parameters given by (1)-(3) become energy dependant.

What happens to gravity in the energy region between $10^{16} - 10^{19} \, GeV$? This has been addressed in refs [23,24] with consequences for the early universe and the big bang singularity.

Recent suggestions as to why gravity is very weak and alternate attempts to address the so called hierarchy problem (i.e., large gap between electroweak and Planck scale) invoke the concept of large extra dimensions. The point made is [25] that (4+n) dimensional gravity would become as strong as other interactions long before $L_{pl}$ (or $E_{pl}$!). If the unification occurs at the electroweak scale, $L_{EW} \approx 10^{-17} \, cm$, corresponding to $TeV$ energies, the size $R_C$ of the compact extra dimensions turns out to be,

$$R_C = \frac{\hbar c}{M_{WC}^2} \left( \frac{M_{pl}}{M_W} \right)^{2/n} \qquad \ldots(24)$$

Where $M_W$ is the weak boson mass, or in length units, Newton's constant G in n dimensions has dimension $L_{EW}^{2+n}$,

i.e., $R_C^n = L_{EW}^n \left(\dfrac{L_{EW}}{L_{pl}}\right)^2 \approx L_{EW}^n \times 10^{32}$ ...(25)

$\left(L_{EW}/L_{pl} \approx 10^{16}\right)$

($n = 1$, i.e., one large extra dimension, already ruled out by solar system tests of gravity, as $R_C \approx 10^{-17} \times 10^{32} = 10^{15} cm$ )

$n = 2$, gives $R_C \approx 0.1 mm$. Thus the intense current interest in looking for sub-millimetre deviations from Newtonian gravity. [26]

Millimetre length scales also arise when considering dark energy. If the Casimir force between two flat plates separated by a distance $d$ is to balance the repulsive background cosmic energy density, $d$ is given by: [27]

$$d \approx \left(\dfrac{8\pi G \hbar c}{240 \pi \Lambda c^4}\right)^{1/4} = \left(\dfrac{1}{30}\dfrac{\hbar G}{c^3 \Lambda}\right)^{1/4} = \left(\dfrac{L_{pl}^2}{\Lambda}\right)^{1/4} \approx 0.1 mm \,!$$

Here $\Lambda$ is the cosmological constant, which the recent supernova legacy survey gives as $\approx 10^{-56} cm^{-2}$ !

Thus a combination of cosmological and Planck scales seems to imply sub-millimetre scales. This maybe related to dark matter particles (eg. axions) of mass $\sim 10^{-4} eV$. [28]

The present author suggested large extra dimensions in connection with strong gravity, i.e., unification of strong and gravitational interactions (both of which are non-abelian) much earlier. [29]

In the *TeV* picture, gravity becomes as strong as the other interactions (i.e. is unified with) at the weak energy scale ~*TeV*, corresponding to a length of $\sim 10^{-17} cm$ ! So that, effects of quantum gravity are now strong at much smaller energy scales (larger length scales!) and can be tested in the LHC (Large Hadron Collider) which collides proton beams of several *TeV's*. LHC is due to go in operation in 2007.

It is even expected that *TeV* mass mini black holes would be profusely produces! [30] (Similar to Hawking mini black holes of Planck mass which could have been produced at the Planck epoch of the early universe). Just as Planck mass black holes spontaneously decay by Hawking radiation in $\sim 10^{-43}$ sec, the *TeV* black holes which could be produced in LHC would decay in nuclear time scales. [31]

Their production and decay could be detected in the next few years! Predictions of quantum gravity (and string theory) could actually be verified, as the scale has now been brought down from Planck mass to *TeV* energy!

As a final remark, it may be pointed out that the Planck scale and the dark energy are likely to be linked. The Planck scale provides the ultraviolet (short-wavelength) cut-off, while the dark energy (or effective cosmological constant $\Lambda$) provides an infrared long-wavelength cut-off. So if all the electron's mass (for eg.) is due to its self energy given by interactions with all the virtual photons, between the above cut-offs, we get [32] the intriguing relation:

$$\alpha \ln\left(\frac{1}{L_{pl}\sqrt{\lambda}}\right) \approx 1 \qquad \ldots(26)$$

Alternately if $\Lambda$ is fixed from cosmology, and $\alpha$ is known accurately, the UV cut-off (relevant to string theories etc.) is given by a relation such as:

$$l_S \approx \frac{1}{e^{1/\alpha}\sqrt{\Lambda}} \qquad \ldots(27)$$

Above relations also imply that a small variations (eg. with epoch) in the dark energy density can give rise to much smaller logarithmic variations in $\alpha$. These are interesting testable effects for both atomic physics and cosmology.